\documentclass[useAMS,fleqn,usenatbib]{mnras}
\usepackage[T1]{fontenc}
\usepackage{ae,aecompl}
\usepackage{bethmacros}
\usepackage{graphicx}
\usepackage{amssymb}

\def\spose#1{\hbox to 0pt{#1\hss}}
\def\lta{\mathrel{\spose{\lower 3pt\hbox{$\mathchar"218$}}
     \raise 2.0pt\hbox{$\mathchar"13C$}}}
\def\gta{\mathrel{\spose{\lower 3pt\hbox{$\mathchar"218$}}
     \raise 2.0pt\hbox{$\mathchar"13E$}}}

\title[Superbubble Driven Outflows]{Cosmological Galaxy
Evolution with Superbubble Feedback I:  Realistic Galaxies
with Moderate Feedback}

\author[Keller et al.]{B.\,W. Keller$^{1}$\thanks{Email: bwkeller `at' mcmaster.ca},  J. Wadsley$^{1}$, H. M. P. Couchman$^1$
\vspace*{6pt}\\
$^1$Department of Physics and Astronomy, McMaster University, Hamilton, Ontario, L8S 4M1, Canada
}

\begin{document}
\maketitle
\label{firstpage}

\begin{abstract}
    We present the first cosmological galaxy evolved using the modern smoothed
    particle hydrodynamics (SPH) code \textsc{GASOLINE2} with superbubble
    feedback.  We show that superbubble-driven galactic outflows powered by Type
    II supernovae alone can produce $\rm{L^*}$ galaxies with flat rotation
    curves with circular velocities $\sim 200\; \rm{km/s}$, low bulge-to-disc
    ratios, and stellar mass fractions that match observed values from high
    redshift to the present.  These features are made possible by the high mass
    loadings generated by the evaporative growth of superbubbles.  Outflows are
    driven extremely effectively at high redshift, expelling gas at early times
    and preventing overproduction of stars before $z=2$.  Centrally
    concentrated gas in previous simulations has often lead to unrealistically
    high bulge to total ratios and strongly peaked rotation curves.  We show that
    supernova-powered superbubbles alone can produce galaxies that agree well
    with observed properties without the need for additional feedback mechanisms
    or increased feedback energy.   We present additional results arising from
    properly modelled hot feedback.
\end{abstract}


\section{Introduction}
The theory of galaxy formation is a cornerstone of modern cosmology.
$\rm{\Lambda CDM}$ accurately predicts the detailed properties of the Cosmic
Microwave Background, and the formation of large-scale cosmic structure.
Understanding how this yields the small-scale properties of the galaxies we see
today and at high redshift requires a good understanding of the
physics at play within these individual galaxies:  star formation, feedback, gas
accretion, etc.

The basic theoretical picture of galaxy formation is now well-established
\citep{Rees1977, White1978}.  The complex details are typically probed through
both simulation and semi-analytic techniques \citep{Kauffmann1999,Bower2006}.
Simulators typically employ large-scale cosmological boxes and zoomed in
simulations, in which regions sliced out of those larger volumes are re-simulated
at higher resolution \citep{Katz1993, Governato2004, Stinson2010, Brooks2011}.
Recent studies \citep{Dekel2006,Woods2014} have shown that the picture of how
gas is fed to galaxies is not simple:  cold flows can be funneled along
filaments in the cosmic web, bypassing the virial shock and directly supplying
gas to the inner galaxy.  Gas accretion and expulsion by stellar feedback is fundamental
to how galaxies evolve over time, determining not only when and how many stars
form, but the kinematic properties of those stars as well.

Stellar feedback has a greater impact than merely regulating star formation by
heating the ISM or increasing its turbulence.   It has long been understood that
the cumulative effects of multiple supernovae can eject gas from the galaxy,
powering a galactic outflow or wind \citep{Mathews1971,Larson1974}.  Galactic
winds are common in high-redshift galaxies, and can be seen in the nearby
Universe in starburst galaxies (See review by
\citealt{Veilleux2005}).  The detection of both blue-shifted rest-frame UV
absorption in observations of high redshift galaxies
\citep{Weiner2009,Steidel2010,Martin2012} along with broadened $H\alpha$
emission lines \citep{Heckman1987,Genzel2011,Newman2012} strongly suggests the
presence of hot, outflowing material surrounding these galaxies.  Galactic winds
appear to be ubiquitous at high-redshift, and are likely a key factor in the
evolution of galaxies in the early Universe.  \citet{Muratov2015} estimated that
mass loadings for winds peaked at roughly $\eta=\frac{\dot M_{wind}}{\dot
M_{*}}\sim10$ at high redshift and that a significant
fraction of ejected material can escape beyond the virial radius.  By ejecting
material from a galaxy's disc, and storing it in the hot, gravitationally bound
circumgalactic medium (CGM), star formation at high redshift can be strongly
regulated, while at the same time providing a reservoir of gas that can be
accreted at late times, ensuring that star formation at low redshift can
continue \citep{Marasco2012}.  How this CGM is created is complex, as it is
potentially pierced by cold flows of infalling material, outflows from the
central galaxy, and the accretion of dwarf satellites.

The first generations of cosmological simulations including baryonic physics
(hydrodynamics, star formation, etc.) suffered from a number of serious
problems.  Nearly every simulation, regardless of halo mass, produced far too
many stars, and tended to form these stars too early \citep{Abadi2003,
Governato2009, Stinson2010, Brooks2011}.  Simulations of disc galaxies produced
bulge dominated galaxies rather than the thin discs we observe.  Together with
the discrepancies seen between large dark-matter only simulations and
observations, it appeared that the standard dark energy + cold dark matter
cosmology $\rm{\Lambda CDM}$ might need to be modified to adequately explain the
properties of both individuals galaxies and populations of galaxies that we see.
\cite{Scannapieco2012} showed these problems were ubiquitous among codes,
leading simulators to consider a role for stronger feedback.  Early feedback
models (e.g. \citealt{Katz1993}) tended to deposit the energy of feedback into a
poorly-resolved region of the interstellar medium, subjecting it to
over-cooling.  Stronger feedback can be achieved by limiting cooling or
increasing the total energy.  It must also ensure that feedback energy couples
strongly to the gas.  This strong coupling doesn't just heat the ISM, or disrupt
only the densest regions, but drives outflows that actively remove gas from the
disc of the galaxy.   The Aquila comparison of 13 different simulation codes and
subgrid physics models \citep{Scannapieco2012} showed that only those cases with
strong outflows could produce galaxies with realistic star formation histories.
The outflow models used were somewhat ad hoc and extremely aggressive.  The
cases that were capable of ejecting sufficient gas from the galaxy disc to
moderate the stellar mass of the halo still failed to produce stellar discs and
largely shut down low redshift star formation.

Galactic winds have been a part of cosmological galaxy simulations for 
some time \citep{Springel2003}, and recent simulations have investigated these
winds in detail.    \citet{AnglesAlcazar2014} showed that these winds can dramatically
alter the star formation history, kinematics, and morphology of galaxies at
redshift 2.  By explicitly creating galactic winds with a variety of
mass-loadings and wind velocities, they showed that strong winds are essential
to producing the gas-rich, extended, and turbulent discs that are typically
observed in high redshift star forming galaxies.  Unfortunately, without 
simulations to $z=0$, interpreting exactly how these high-redshift winds impact
present-day galaxies is difficult. If outflowing material falls back onto the
galactic disc within a Hubble time, the effects of high-redshift winds may be
seen in the form of increased star formation and inflows at low redshift.  

A key question remains: What processes set the mass loading and velocity of
these winds?  Early work tied galactic outflows to the SNe energy
\citep{Springel2003}.  Some studies (e.g. \citealt{Murray2005,Krumholz2013})
have suggested that radiation pressure is needed to drive sufficient
galaxy-scale winds.  Others argue that galactic winds could be powered by cosmic
ray buoyancy \citep{Ipavich1975,Breitschwerdt1991,Socrates2008}.

Today, multiple different models of strong feedback have begun to produce
galaxies with the correct number of stars \citep{Aumer2013}, reasonable star
formation histories \citep{Stinson2013,Agertz2014,Munshi2013}, and correct
morphologies \citep{Guedes2011,Brook2012,Christensen2014}.   Unfortunately, many
of these successes have come at the cost of increasing complexity in star
formation and feedback methods, crude assumptions regarding the physics of the
feedback-heated gas and somewhat arbitrary increases in the feedback energy per
unit mass in stars.  

In many cases, strong feedback simply means more energy.  Many modern feedback
models augment the energy input from Type II supernovae ($\sim10^{51}\;
\rm{erg}$ per star above $8\; \rm{M_\odot}$), with that arising from stellar
winds, UV ionization, supermassive black holes (SMBH), or radiation pressure
\citep{Vogelsberger2013, Agertz2014}.  Feedback models such as
\citet{Stinson2013} group these sources of energy as early stellar feedback.
Since the first supernovae occur $\sim4\; \rm{Myr}$ after star formation, these
feedback mechanisms begin depositing feedback before SN-only methods would.
Unfortunately, how much energy from these early feedback effects actually
couples to the surrounding ISM rather than radiating away is highly uncertain.
In fact, even the coupling of comparatively simpler SN feedback still is a
matter of some debate.  Many new feedback models \citep{Agertz2013, Aumer2013,
Hopkins2013} treat each of these feedback mechanisms explicitly, modelling the
input of energy and momentum from each component separately.  

In addition to increasing the total amount of energy deposited by stellar
feedback, these models often also include components designed to prevent energy
from feedback radiating away (a problem discussed thoroughly in
\citealt{Thacker2000}).   The energy can be prevented from cooling completely
for a while \citep{Stinson2013} or it can be initially placed into a non-cooling
reservoir that leaks back into regular thermal energy \citep{Agertz2013}.
Alternately, depositing thermal feedback into a sufficiently small mass, allows
it to always heat gas to the same high temperature where cooling times are long
\citep{DallaVecchia2012}.  Depositing feedback as kinetic avoids initial
radiative losses \citet{Springel2003,Agertz2013,Hopkins2013}.  

While these techniques do help solve the overcooling problem, they all come with
some drawbacks.  Fixed-temperature thermal feedback is stochastic, and requires
the additional free parameter of the feedback temperature. Cooling shutoffs
completely disable radiative losses, where in nature these losses are suppressed
in some regions but can remain strong in others, depending on the structure of
the ISM and the clustering of stars.  Kinetic feedback is almost always paired
with a temporary decoupling of hydrodynamic forces on feedback-accelerated gas.
This is necessary to prevent this gas from shock-heating and potentially
reintroducing the overcooling problem (as shown by
\citealt{Creasey2011,Durier2012}).  Decoupling allows winds to escape.  However,
this makes it impossible to study the detailed behavior of these winds, and how
they interact with the ISM.  This makes the mass loading an imposed parameter.

Of these methods, \citet{DallaVecchia2012} showed the interesting result that,
for supernovae alone, depositing feedback energy into a pre-specified amount of
mass, without any cooling shutoffs, can give reasonable star formation rates and
strong galactic outflows in isolated galaxies.  \citet{Keller2014} argued that
what sets this mass is thermal conduction in superbubbles, and used that
mechanism to build a new way of simulating supernovae feedback that lacked the
resolution dependence, additional complexity, and ad-hoc additions of many
current feedback models.  By focusing on superbubbles, and the evaporation of
cold gas to determine mass loading, superbubble feedback gives realistic gas
behavior, and is effective at both regulating star formation and driving
galactic outflows without introducing free parameters.

The original McMaster Unbiased Galaxy Simulations (MUGS; \citealt{Stinson2010})
showed that with stellar feedback, the observed color-magnitude relationship and
Tully-Fisher relation could be produced in simulated $\rm{L^*}$ galaxies.  It
failed, however, to produce galaxies with the correct stellar mass fraction and
star formation history, overproducing stars over the entire cosmic history, and
grossly overproducing them at high redshift.  It also produced galaxies with
bulge-to-disc fractions larger than those seen in nature, and with sharply
peaked rotation curves.  \citet{Stinson2013} showed that the addition of early
stellar feedback could alleviate most of these problems.  That model does not
take into account the potentially complex coupling of stellar winds or radiation
pressure to the surrounding ISM.  Instead, a substantial (fixed) fraction of the
stellar bolometric luminosity was applied as feedback heating.

In this paper, we begin first with an overview of the simulation methods used,
both gas microphysics and the star formation and feedback techniques.  We then
present the results of a suite of 4 simulations using an initial
condition from the MUGS sample, each generated with different stellar feedback
models or energetics.  The resulting galaxy properties,  as well as the halo
evolution are examined over the lifetime of the galaxy.  Finally, we discuss how
the superbubble-driven outflows change with time, and how they ultimately result
in a realistic galaxy at the present epoch.

\section{Methods}

These simulations were run using a modern update to the SPH code {\sc GASOLINE}
\citep{Wadsley2004}, {\sc GASOLINE2}.  The changes in this new code include a
sub-grid model for turbulent mixing of metals and energy \citep{Shen2010}, and a
modified pressure force form similar to that proposed by \citet{Ritchie2001}
which is functionally equivalent to \cite{Hopkins2013}.  These changes solve the
problems seen in Kelvin-Helmholtz and blob destruction tests with SPH
\citep{Agertz2007}.  These and other features are discussed in
\citet{Keller2014}.  Accurately modelling mixing in multiphase gas is essential
for accurately simulating the ISM and CGM.

\subsection{Simulations}
For this initial study, we have selected the initial conditions from one of the
original MUGS galaxies. We selected an intermediate-mass halo, g1536, allowing
us to compare to a number of other studies that have examined this particular
halo (e.g. \citealt{Stinson2013,Woods2014}).  At $z=0$ this halo has a virial
mass of $8\times10^{11}\; \rm{M_\odot}$ and a spin parameter of 0.017.  It had
its last major merger at $z=2.9$.  We have a gas mass resolution of
$M_{gas}=2.2\times10^5\; \rm{M_{\odot}}$, and use a gravitational softening
length of $\epsilon=312.5\; \rm{pc}$.  The details of how this IC was created
can be found in \citet{Stinson2010}.  We choose to focus on a single galaxy for
this paper as it allows us to make a direct comparison of the impact feedback
physics makes with relatively little expense.  Naturally, this limits our
ability to comment on population-wide effects.  We leave this discussion
for a forthcoming paper in which we introduce 17 additional $L^*$ galaxies.

We compare four test cases using the same initial conditions: no stellar
feedback (an absolute lower bound for looking at the effects of feedback);
superbubble feedback (our fiducial case); blastwave feedback (the feedback
method used in \citet{Stinson2010}, first described in \citealt{Stinson2006}); and
superbubble feedback with double the standard feedback energy ($2\times10^{51}\;
\rm{erg}$ per SN).  The high feedback energy case uses more feedback than is
predicted by \citet{Leitherer1999}, but is within the range of feedback energies
currently being used in cosmological simulations \citep{Schaye2015, Agertz2014,
Vogelsberger2013}.  

Halos were found in each of the simulations using the Amiga Halo Finder
\citep[AHF;][]{Knollmann2009}.

\subsection{Star Formation and Feedback}
Our suite of simulations use a range of different feedback
processes.  For all of the simulations shown in this paper, we use a common
star formation prescription.  Stars are formed at a rate proportional to the
local free-fall time of gas, such that
\begin{equation}
    \dot \rho_* = c_* \frac{\rho_{gas}}{t_{ff}}
    \label{sflaw}
\end{equation}
For each of these simulations, we used an efficiency parameter $c_* = 0.1$, the
value used by \citet{Stinson2013}.  Stars are allowed to form in a converging
flow when gas is cooler than $1.5\times10^4\; \rm{K}$, and with a density set to
that allowed by the gravity resolution, $\rho = M_{gas}/\epsilon^3 = 9.3\;
\rm{cm^{-3}}$.

The amount of supernova feedback per unit stellar mass is determined using a
\citet{Chabrier2003} IMF.  With $\sim10^{51}\; \rm{erg}$ per supernova this
gives $\sim10^{49}\; \rm{erg\;M^{-1}_{\odot}}$.   A notable difference between
this simulation and those of \citet{Stinson2013} and others is that we do not
include early feedback, processes such as stellar winds and radiation that can
inject energy before supernovae occur $\sim 4\; \rm{Myr}$ after the first
massive stars form.  A primary role for early stellar feedback is to disrupt the
densest molecular gas.  In these simulations, this dense molecular gas cannot be
resolved, never being formed (and thus it does not form or need to be
destroyed).

The feedback recipe used in our main simulations is the superbubble method
presented in \citet{Keller2014}.  This method deposits feedback into resolution
elements in a brief multi-phase state.  These particles each have separate
specific energies, masses and densities, for the hot bubble and surrounding cold
ISM, which includes the swept up shell.  This allows the method to calculate a
separate density and cooling time with the respective densities for each phase,
rather than the average density of both phases.  Multiphase particles are also
prevented from forming stars: the average temperature of the two phases is
essentially always above our temperature threshold for star formation.  More
details on this method can be found in \citet{Keller2014}.

The addition of thermal conduction and evaporation introduces some additional
time step constraints in order to ensure the stability of integration.  However,
since the thermal conduction rate is capped by the saturation imposed by the
electron soundspeed, this time step is at worst 1/17 the Courant time step. In
fact, we see an overall speedup when using superbubble vs. blastwave feedback,
as gas can no longer exist in the regime of high density and temperature (and
thus small Courant times).  The average computation time per step was $\sim
25\%$ faster using superbubble feedback compared to blastwave feedback.  These
benefits are dominant late in the run, as the total amount of dense gas becomes
larger. We do see some slight increased cost early in the run, with a slowdown
of $\sim50\%$ before z=4.  The additional time-step constraints here are similar
to the time-step constraints required for other methods as well, such as
decoupled kinetic winds \citep{Springel2003}.

\subsection{Gas Cooling and Physics}
We adopt the same gas cooling physics as a number of past simulations using
\textsc{GASOLINE} \citep{Stinson2013,Keller2014}.  The method for cooling we use
here was originally presented in \citet{Shen2010}.  Our simulations use cooling
rates from \textsc{CLOUDY} \citep{Ferland2013}, and  include a
redshift-dependent UV background, Compton cooling, and primordial and metal
cooling from 10 to $10^9\; \rm{K}$.  This sets these simulations apart from many
past simulations \citep{Governato2009,Brook2011,Guedes2011}, which did not
include high temperature metal cooling.  We impose an artificial pressure floor,
using a method described by \citet{Robertson2008} to prevent spurious Jeans
fragmentation (as cold, dense gas has both Jeans length and Jeans mass below the
resolution limit of our simulation).  We also enforce that the minimum SPH
smoothing length a particle may have is $\epsilon/4$.  This is equivalent to a
density floor of $400\; \rm{cm^{-3}}$ (note that for two-phase superbubble
particles, this is the maximum \textit{mean} density).  These are comparable to
the parameter choices in another recent re-simulation of the MUGS initial
conditions, \citet{Stinson2013}.

\begin{figure*}
    
    \includegraphics[width=\textwidth]{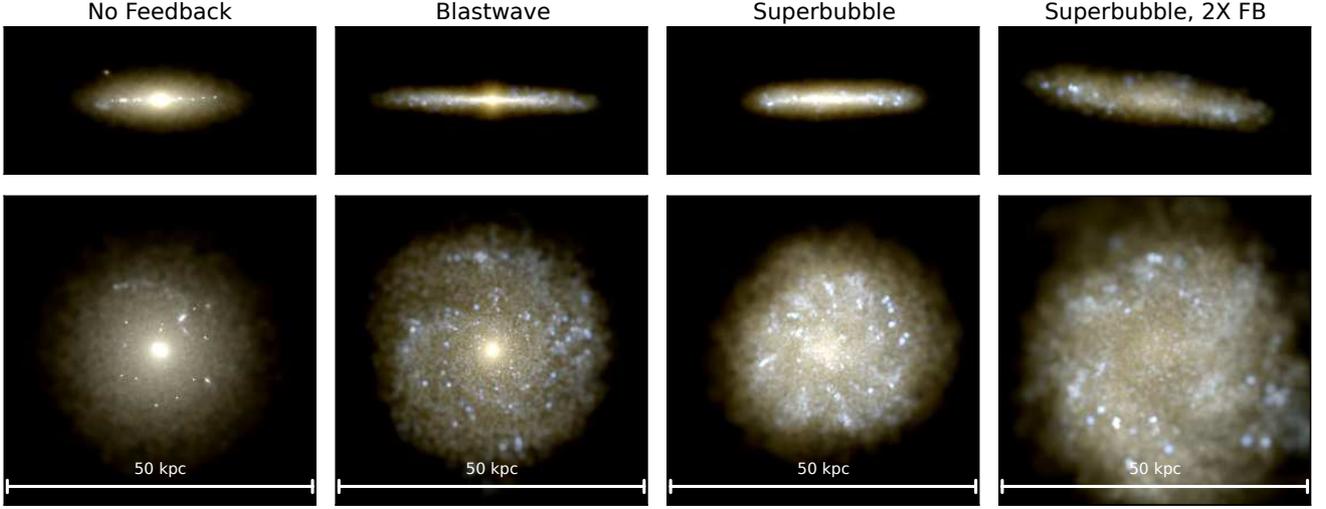}
    \caption{Mock stellar image of each galaxy at $z=0$.  The top row
    shows the galaxies edge-on, while the bottom shows them face-on.  The
    no-feedback case shows very little disc, while the blastwave feedback case has
    a thin disc with a prominent bulge.  The superbubble galaxy appears nearly
    bulgeless, composed only of a truncated disc. Note also that the superbubble
    galaxy is bluer than the blastwave or no-feedback galaxies, evidence of a
    younger stellar population.}
    \label{stellar_image}
\end{figure*}
\begin{figure*}
    \includegraphics[width=\textwidth]{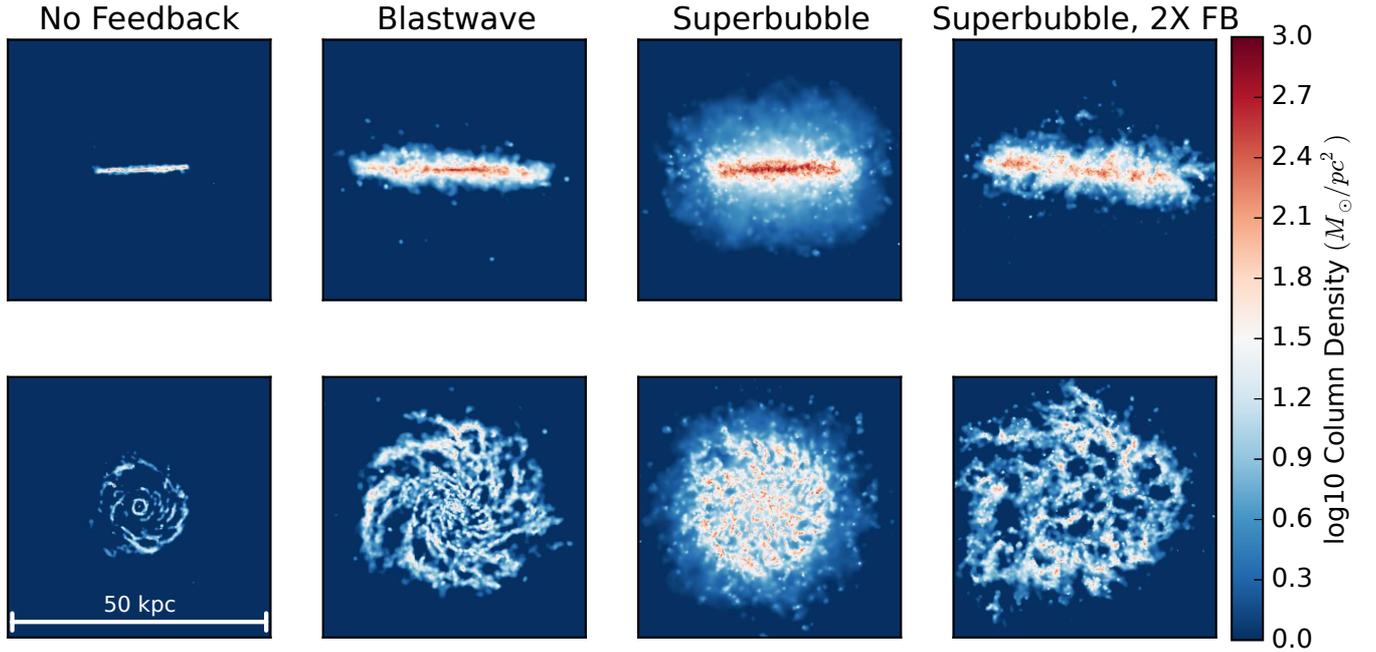}
    \caption{HI column density for the four test cases at z=0.  The no feedback case has
    exhausted nearly all of the gas within the disc, leaving only diffuse
    wisps.  The two superbubble cases (especially with doubled supernova energy)
    have lofted a large amount of gas out of the disc, and some entrained dense
    clumps can be seen in this outflowing material.  The superbubble gas discs
    are much more flocculent than the blastwave disc, and where the feedback
    energy is doubled, the disc is strongly disrupted.}
    \label{column_density}
\end{figure*}
\begin{figure}
    \includegraphics[width=\columnwidth]{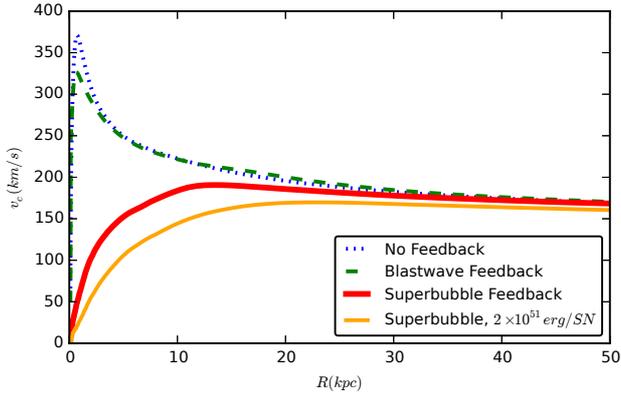}
    \caption{Rotation curves for each of our test cases.  As is clear,
        the superbubble cases have much lower central concentrations, giving a
        a rotation curve that rises to a flat $200\; \rm{km/s}$.  The peaked rotation
        curves in the cases with blastwave or no feedback are a result of their
        failure to remove low angular momentum gas at high redshift, giving
        bulge-dominated, centrally concentrated galaxies.}
    \label{rotation_curve}
\end{figure}
\begin{figure}
    \includegraphics[width=\columnwidth]{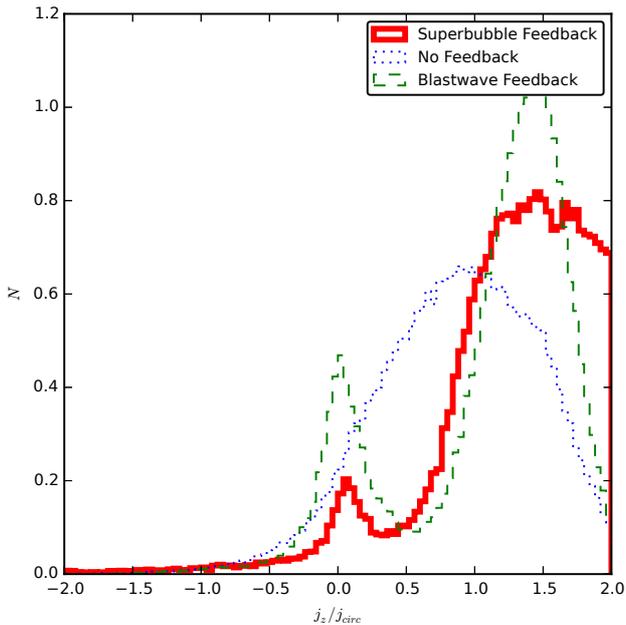}
    \caption{Histogram of stellar orbit circularity.  The bimodal distribution
        shows the bulge component, with a circularity near 0, and the disc
        component, with a circularity near 1.  Note that without feedback, this
        bimodality disappears, as the galaxy morphology becomes spheroidal.  The
        values here are normalized so that each curve has a total integral of
        unity.  As is clear, very few stars produced in a galaxy with superbubble
        feedback have low angular momentum, giving us a disc-dominated galaxy.}
    \label{circularity}
\end{figure}
\begin{figure}
    \includegraphics[width=\columnwidth]{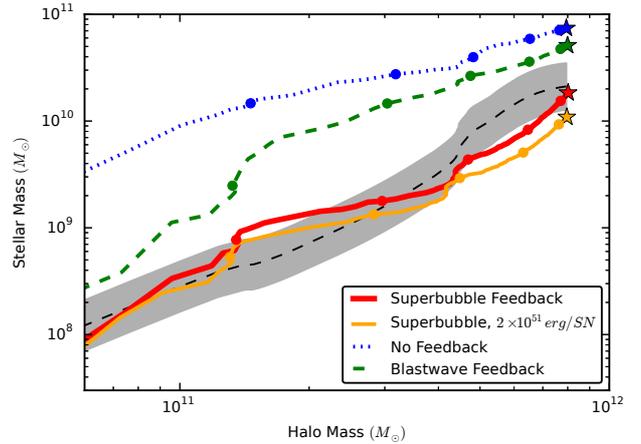}
    \caption{Stellar mass growth as a function of the total halo mass.  The grey
    band and black dashed curve show mean values and uncertainties from the
    abundance matching results of \citet{Behroozi2013}.  The points show values at
    redshifts 3,2,1,0.5, and 0.1.  The stars show the final values at redshift 0.
    For essentially the entire evolution of the halo, both the no feedback galaxy
    and the blastwave galaxy overproduce stars.}
    \label{mass_growth}
\end{figure}
\begin{figure}
    \includegraphics[width=\columnwidth]{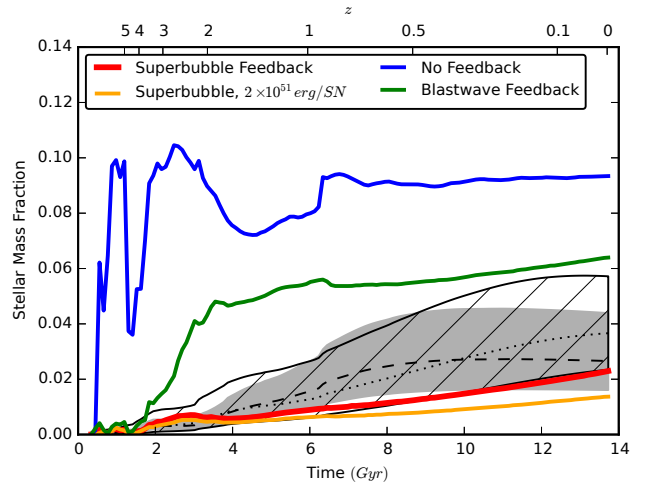}
    \caption{Stellar mass fraction as a function of time/redshift.  As in
    figure~\ref{mass_growth}, only the fiducial superbubble run produces stellar
    mass fractions within observed uncertainties for the entire history of the
    halo. The grey band is as in figure~\ref{mass_growth}, and here we also show, in
    the hatched region with the dotted curve, values and uncertainties from
    \citet{Moster2013}.  The dip that can be observed in the no feedback case
    at high redshift come from the growth of the halo, as it accretes dark
    matter dominated dwarves.}
    \label{stellar_fraction}
\end{figure}
\begin{figure}
    \includegraphics[width=\columnwidth]{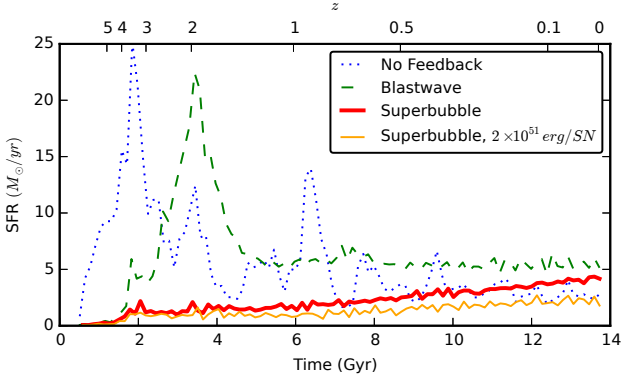}
    \caption{Star formation rates for all 4 simulations.  The low star formation
    rate seen after $z=1$ in the no feedback case is due simply to the high z
    star formation consuming most of the available gas.}
    \label{sfr}
\end{figure}
\begin{figure}
    \includegraphics[width=\columnwidth]{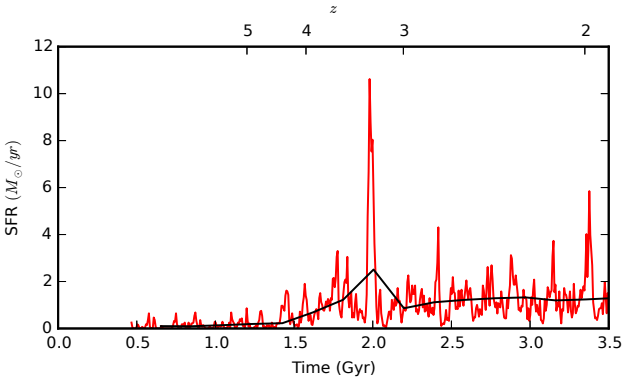}
    \caption{Star formation rate up to $z=2$ for our superbubble galaxy.  As is
        clear, when sampled on short ($7\; \rm{Myr}$) timescales, the star
        formation rate in the superbubble galaxy is characterized by strong
        bursts, despite its quiescent behavior averaged over longer
    timescales.}
    \label{burstiness}
\end{figure}
\begin{figure}
    \includegraphics[width=\columnwidth]{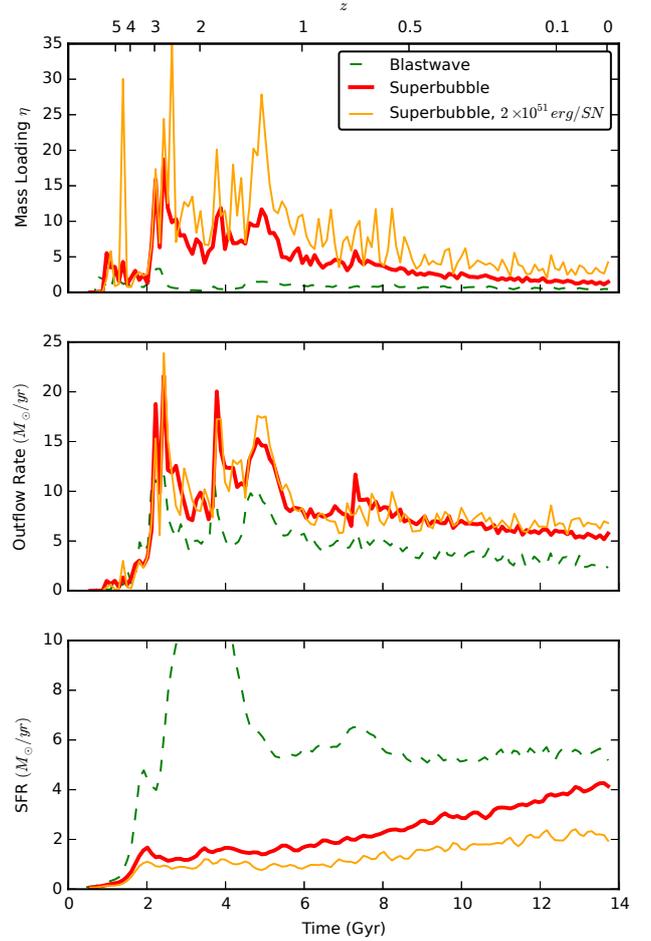}
    \caption{Galactic wind mass-loading and outflow rates for each feedback
    model.  As is clear, superbubble feedback powers winds with significantly
    higher mass-loadings, and, despite forming less than half as many stars the
    blastwave galaxy, results in more total outflowing gas.  This is key to the
    regulatory power of superbubbles.  Despite the peak in outflow rate in the 
    blastwave case between $z=4-1.5$, the massively increased star formation rate
    seen in the the smoothed SFR during this time (bottom plot) means that the
    mass-loadings never go above 5, and baryon expulsion is inefficient.}
    \label{massloading}
\end{figure}
\begin{figure}
    \includegraphics[width=\columnwidth]{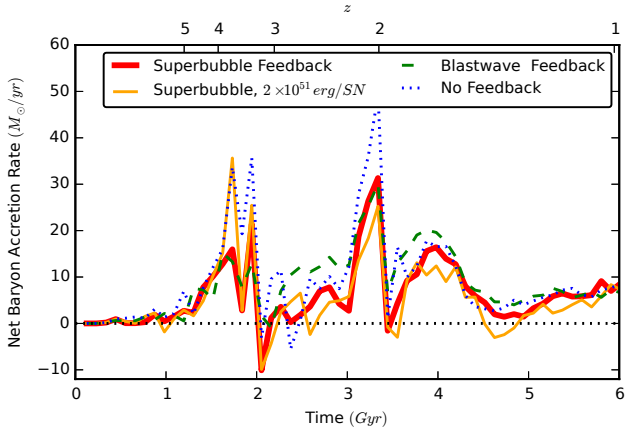}
    \caption{Net baryonic accretion (stars and gas) for each test case.  It is
        clear that superbubble feedback expels baryons much more effectively
        than the blastwave feedback model (although even weak feedback does help
        somewhat compared to no feedback whatsoever).}
    \label{net_accretion}
\end{figure}
\begin{figure}
    \includegraphics[width=\columnwidth]{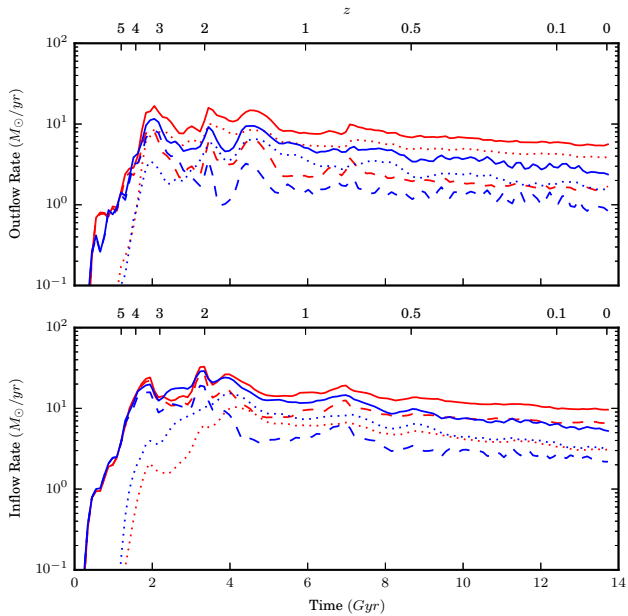}
    \caption{Inflow and outflow rates for superbubble (red) and blastwave
        (blue) run.  The solid lines show
        the total inflow rate, while the dashed and dotted lines show rates for
    cold and hot gas (above/below $10^5\; \rm{K}$) respectively.  }
    \label{inflow_outflow}
\end{figure}
\begin{figure}
    \includegraphics[width=\columnwidth]{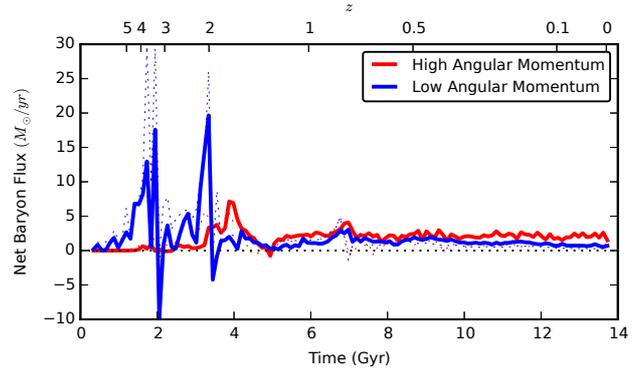}
    \caption{In all four test cases, including the fiducial superbubble run
        shown above by the solid curve, and the no-feedback case shown by the
        dotted curve, the vast majority of low angular momentum material, with
        specific angular momentum $ j_z < 500\; \rm {kpc\; km/s}$, is accreted
        by $z=2$.  The accretion of high angular momentum material, with $j_z >
        1500\; \rm{kpc\; km/s}$ rises to peak at $z=2$, and continues to accrete
        $\sim 2\; \rm{M_\odot/yr}$ at $z=0$.  As can be seen here, the net flux
        of low angular momentum material at high redshift is suppressed, while
        the accretion of high angular momentum material at low redshift is
    actually {\it enhanced} by superbubble feedback.}
        \label{angular_momentum_netflux}
\end{figure}
\begin{figure}
    \includegraphics[width=\columnwidth]{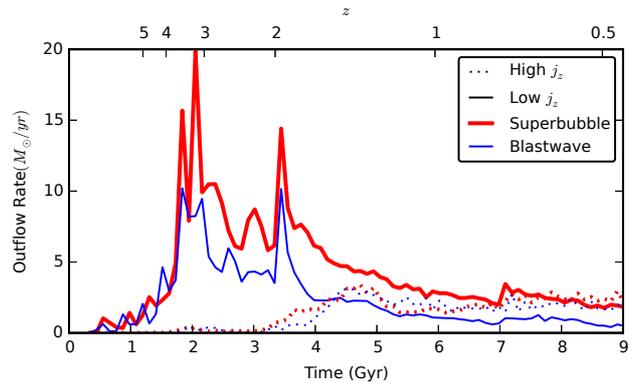}
    \caption{High-redshift winds preferentially remove low angular momentum
        material.  As can be seen here, significantly more low angular momentum
        material is removed by superbubble feedback than by the weaker
        blastwave feedback, but the amount of ejected high-angular
        momentum material is roughly equal.}
        \label{angular_momentum_outflow}
\end{figure}
\begin{figure}
    \includegraphics[width=\columnwidth]{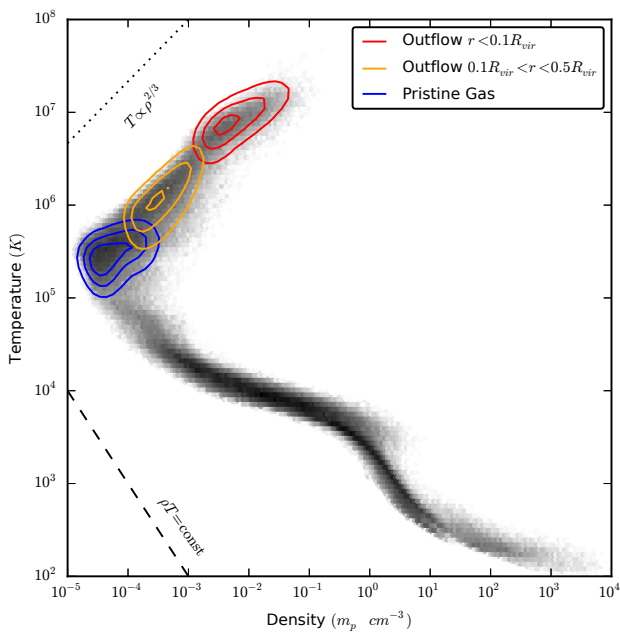}
    \caption{Phase diagram of gas in the halo at $z=0$.  The hot, coronal gas
        with temperatures above $10^5\;K$ contains roughly 40\% of the gas mass
        in the halo.  This gas is a mixture of virial-shocked pristine gas,
        which has never been within the interior $0.1R_{vir}$, and outflowing
        material ejected from the galaxy.  This buoyant, high-entropy gas exits
        the galaxy at high temperature, and cools adiabatically as it rises
        through the CGM.  The three colored contours show the major components
        of this hot coronal gas.  Pristine gas, never falling within the
        interior, is shown in blue.  Young superbubbles, not yet having broken out
        of the galaxy, are shown in red.  Outflowing material that was once inside the
        galaxy but is now cooling adiabatically as it rises through the CGM can be seen
        in orange.  The contours show the 1.5\%, 1\%, and 0.5\% levels for the
        total mass in each cut.}
        \label{phase}
\end{figure}

\section{Results}
\begin{table*}
    \begin{tabular}{ c c c c c c c}
        \hline
        Simulation & $M_{vir}$ & $M_{gas}$ &
        $M_{star}$  &$M_{bary,0.1vir}$ & $M_{bulge}$ & $M_{disc}$ \\
        \hline
        No Feedback & $7.94\times10^{11}$ & $6.21\times10^{10}$ &
        $7.42\times10^{10}$ & $8.09\times10^{10}$ & $9.34\times10^{9}$ &
        $2.88\times10^{10}$\\
        Blastwave Feedback & $7.99\times10^{11}$ & $8.05\times10^{10}$ &
        $5.11\times10^{10}$ & $7.94\times10^{10}$ & $8.09\times10^{9}$ &
        $1.75\times10^{10}$ \\
        Superbubble Feedback & $8.02\times10^{11}$ & $1.19\times10^{11}$ &
        $1.84\times10^{10}$ & $6.25\times10^{10}$ & $1.14\times10^{9}$ &
        $1.15\times10^{10}$ \\
        Superbubble, $2\times10^{51} erg/SN$ & $7.96\times10^{11}$ &
        $1.20\times10^{11}$ & $1.09\times10^{10}$ &  $4.92\times10^{10}$ & 
        $5.96\times10^{8}$ & $6.98\times10^9$\\
    \end{tabular}
    \caption{Halo components at $z=0$. $M_{disc}$ and $M_{bulge}$ are the
    determined by the angular momentum of the stars (as detailed below).  All
    masses are in $M_\odot$.}
    \label{properties}
\end{table*}

\subsection{Redshift zero Disc Properties}
As can be seen in Figure~\ref{stellar_image}, the galaxy produced with
superbubble feedback is disc-dominated and blue.  The stellar disc in the case
without feedback is nearly non-existent, with an ellipsoidal morphology.  The
superbubble disc appears somewhat thicker and truncated compared to the disc
produced with blastwave feedback.  The truncation can be seen in the reduced
stellar scale length of the superbubble galaxy, $2.9\;\rm{kpc}$ vs.
$3.8\;\rm{kpc}$ for the superbubble vs. blastwave galaxies.  The thickening
appears to be quantitatively insignificant, however, as both discs have stellar
scale heights of $0.9\;\rm{kpc}$ This may be due to the the disc being disrupted
at large radii, where the surface densities are lower, allowing superbubbles to
grow larger and escape more easily.  The increased feedback case is
interestingly somewhat tilted compared to the rotation of the halo as a whole.  

The gas component of this galaxy can be seen in Figure~\ref{column_density}.
The superbubble feedback is clearly much more effective at ejecting gas from the
disc, creating a halo of clumpy HI gas.  In the doubled feedback case, the disc
is heavily disrupted, with no spiral structure visible in either the stellar or
gas density images.  Thus the larger stellar disc is directly related to a more
extended gaseous disk.

The rotation curve in Figure~\ref{rotation_curve} shows that this ejection also
gives a flattened rotation curve without the central peak seen in simulations
with blastwave or no feedback.  This tells us that the mass distribution is much
less centrally concentrated when superbubble feedback is used, more evidence
that superbubble feedback is effective in preventing bulge formation.

One of the primary problems in older simulations was a much larger
bulge-to-total fraction based on a kinematic decomposition.  We adopt the same
kinematic decomposition as the original MUGS study.  We calculate, for each star
within the halo, a circularity parameter, which is simply the ratio of the
specific angular momentum component perpendicular to the disc ($j_z$) and the
specific angular momentum for a perfect circular orbit in the same potential
that the star sees $j_{circ}$.  As in the original MUGS study, we identify bulge
stars as having $j_z/j_{circ} < 0.7$ and orbital radii of $< 5\; \rm{kpc}$.  We
find, using these criteria, that our fiducial B/T ratio at $z=0$ is 0.09,
slightly smaller than the Milky Way's $\sim0.14$, and greatly reduced compared
to the 0.55 found in the original MUGS paper, and well within the observational
constraints from \citet{Allen2006}.  As these numbers, and the distribution of
this circularity parameter found in Figure~\ref{circularity} show, the vigorous
expulsion of central gas with superbubble feedback is a powerful bulge
prevention and destruction mechanism.  Each of the 13 samples from the Aquila
project \citep{Scannapieco2012} suffered from either from a bulge-dominated
stellar circularity profile, or from a massively peaked (or in some cases simply
too high) rotation curve.  Even the three cases that managed to produce a
reasonable stellar mass fraction (each of these generated strong outflows,
either through SMBH feedback or wind decoupling) still failed to produce
disc-dominated galaxies.  Of their sample, only four cases had more than 40\% of
the stellar mass in disc stars, but all of these cases massively overproduced
stars.

The ejection of gas is evident if we look at the baryon content of the interior
part of the halo, where the disc resides, in this case, simply material within
$0.1R_{vir}$,.  With superbubble feedback, the baryon fraction of this inner
region is 0.30, reduced from 0.37 without feedback, and 0.35 with blastwave
feedback.  This means that nearly 20\% of the baryons that would be available to
form stars have been blown out of the galaxy disc, roughly twice the amount that
was removed by the older feedback model.  The baryon deficit in the superbubble
is comparable to the total stellar mass of the galaxy (the fiducial case has a
$1.5\times 10^{10}M_\odot$ deficit in baryons (compared to the no-feedback
case), and a total stellar mass of $1.8\times 10^{10}M_\odot$).  

Of what remain, only 29\% of the baryonic mass in the superbubble galaxy disc is
in stars, compared to 63\% in the blastwave galaxy, and 89\% in the no
feedback galaxy.  The basic properties of each halo can be found in
table~\ref{properties}.

\subsection{Halo Evolution and Star Formation}
Ejecting baryons from the disc is essential to producing galaxies that fit the
$M_*/M_{halo}$ relationship predicted by \citet{Behroozi2013},
\citet{Moster2013} and others.  As Figures~\ref{mass_growth}
and~\ref{stellar_fraction} show, the galaxies either lacking feedback or using
blastwave feedback alone fail to regulate star formation, diverging from the
expected abundance matching relation at $z\sim3$ for the blastwave, and before
$z=5$ without feedback, giving galaxies that lie above the abundance-matching
relationships over nearly their entire history and mass evolution.  With
superbubble feedback, this galaxy lies within the range of observed stellar
masses over its entire evolution, although on the low side of this range at low
redshift.  Arbitrarily increasing feedback energies, despite having reasonable
star formation rates near $z=0$, under produces stars for most cosmic history,
giving stellar masses {\it lower} than predicted by abundance matching.  Stellar
and total masses were calculated for the region inside $R_{vir}$.  As the halo
grows over time, brief reductions in the stellar mass fraction can be seen for
the no-feedback and blastwave feedback halos in Figure~\ref{stellar_fraction},
simply by the accretion of gas and dark matter that has not yet made it to the
galaxy disc.  Once these mergers complete, the stellar fraction rate leaps up
once again because of the new supply of gas.  The relative flatness in this
figure at low redshift is reflected in the relative flatness in the star
formation rate, shown in figure~\ref{sfr}.  This is simply a side effect of the
massive star formation rates at high redshift:  the bulk of star forming gas has
been consumed, and thus star formation has slowed.

The star formation rate, in bins of $150\: \rm{Myr}$, is shown in
Figure~\ref{sfr}.  With no feedback, or blastwave feedback, star formation is
extremely vigorous before $z=2$, slowing at low redshift as gas available for
star formation is exhausted.  With superbubble feedback, star formation is
relatively constant over nearly the entire history of the halo, with only a
gradual increase towards $z=0$.  This apparent quiescence is a function of the
large time window over which we are averaging our star formation rate.  In
Figure~\ref{burstiness}, you can see clearly that despite having an average star
formation rate of $\sim1\; \rm{M_\odot yr^{-1}}$, this is punctuated by bursts
of star formation of as much as $\sim10\; \rm{M_\odot yr^{-1}}$.  Very similar
results were seen in \citet{Muratov2015}, where bursts of star formation are
followed by peaks in the outflow rates.  This burstiness is a important if
stellar feedback is to drive galactic winds:  to generate large superbubbles,
star formation must be clustered in both space and time.  Whether these bursts
of star formation are able to effectively remove baryons from the disc
ultimately depends on the mass loading of the winds that they drive.

\subsection{Outflow Analysis}
Star formation in the disc drives hot, fast-moving outflows from the central star
forming regions.  These outflows have temperatures of $\sim2\times10^7\; \rm{K}$
as they leave the disc, and also entrain some cold material with them.
Typically outflow velocities are a few hundred $\rm{km/s}$, less than the
escape velocity of the halo, but sufficient to propel gas to large radii before
it begins to fall in again.  

We calculated inflow and outflow rates and velocities by examining particles
within a spherical shell, with inner radius $0.1R_{vir}$ and outer radius
$R_{vir}$ (giving a shell of thickness $0.9R_{vir}$).  We use the $\bar\rho =
200\rho_{crit}$ definition for $R_{vir}$.  Particles with $v_r < 0$ are
inflowing, while those with $v_r > 0$ are outflowing.
Within this shell, the mass flux is determined simply as:
\begin{equation}
    \dot M = \sum_{r_i\in\mathrm{shell}}\frac{M_i v_i}{0.9R_{vir}}
    \label{halo_massflux}
\end{equation}

As figure~\ref{massloading} shows, the outflow behavior of this galaxy is
fundamentally different when superbubble feedback is used.  The wind mass
loading factor $\eta = \dot M_{outflow} / \dot M_*$, a measure of how
efficiently feedback is generating outflows, is roughly an order of magnitude
larger during the bursts of star formation from $z=4-2$, expelling a large
amount of gas from the disc early on.  The resulting suppression of
high-redshift star formation is key to obtaining the correct stellar mass
relation as the halo grows, is unmistakable in figure~\ref{mass_growth}
and~\ref{stellar_fraction}.  Evidence of these outflows can be seen in the $z=0$
column density images in figure~\ref{column_density}.  High-latitude neutral
hydrogen from both ejected and infalling material can be seen in the two
superbubble cases, but is totally absent otherwise.  The smaller amount seen
around the high feedback energy case is simply due to this gas being expelled
further.  The reason superbubbles are so effective at removing mass from the
galaxy and thus regulating the stellar mass depends both on the higher outflow
rates as well as the lower star formation rates needed to drive these outflows,
leading to much larger mass loading factors than the blastwave model can
achieve.  This is clearly shown in figure~\ref{massloading}, as the top panel is
essentially the middle panel divided by the bottom panel.  The outflow rates
driven by superbubbles is at most $\sim2$ times greater than the blastwave model,
but since this is driven by a comparatively tiny amount of star formation, the
massloading, $\eta$, is nearly 10 times greater at high redshift.  The factor of
$\sim2$ higher outflow rate is what we would expect from the roughly $\sim2$
larger baryon depletion from the superbubble vs. blastwave cases.

The effectiveness of the superbubble-driven galactic winds can be clearly seen
in figure~\ref{net_accretion}, where the net accretion of stars and gas is
greatly reduced by the use of superbubble feedback, resulting at $z=1$ in a net
reduction in the total baryonic mass inside $0.5R_{vir}$ of $\sim14\%$ vs.
blastwave feedback, and $\sim30\%$ vs. no feedback whatsoever (along with the 
removal of baryons from the disc discussed earlier. As the outflow
rate drops towards $z=0$, much of this ejected material falls back onto the
disc, and the outflows transition to a fountain mode, actually increasing
the inflow rate seen compared to the blastwave feedback case in
figure~\ref{inflow_outflow}.  This increase in the accretion of gas fuels an
equivalent increase in star formation, as is seen in figures~\ref{sfr}
and~\ref{massloading}.

Figure~\ref{angular_momentum_netflux} shows that the total accreted gas switches
from being dominated by low angular momentum material to being primarily high
angular momentum material at $z\approx2$.  Because $\eta$ is high ($\sim 10$)
during this period, low-angular momentum material is preferentially ejected from
the galaxy without forming a significant number of stars.  These results agree
quite well with the results of \citet{Muratov2015}, which also found $L^*$
progenitors at $z>2$ had wind mass loadings $\eta\sim 10$. The preferential
ejection of low angular momentum gas is clear in
figure~\ref{angular_momentum_outflow}.  The small amount of high angular
momentum material accreted before $z=1$ is ejected at an essentially equal rate
with or without superbubble-driven winds, but low angular momentum material is
propelled into winds at roughly twice the rate when superbubble feedback is
taken into account.

The phase diagram in figure~\ref{phase} shows that, as expected, the galaxy
contains both a hot halo (in which nearly half of the gas mass can be found),
and contains no gas in regions of short cooling time.  This behavior was shown
previously in the isolated galaxy simulations of \citet{Keller2014}.  The hot
halo gas can be divided into three major components. The coolest component is
pristine, virial shocked material that has not ever been accreted (never passed
within $0.1R_{vir}$ of the halo center).  The hottest gas is actually not yet in
the corona, but is the interior of young superbubbles still within the galaxy
disc.  As this material leaves the disc, it cools adiabatically in the lower
pressure environment of the halo, as can be seen in from the $\rho^{2/3}$
adiabatic path it takes through the phase diagram.  Detailed analysis of this
halo gas, especially in comparison to observations like those of
\citet{Steidel2010}, will be explored in a future study of a larger sample of
L* galaxies.

\section{Discussion}

Past work, especially the Aquila comparison \citep{Scannapieco2012}, has shown
that feedback mechanisms which do not remove a large fraction of baryons from
galaxy discs fail to produce realistic spiral galaxies.  Galaxies simulated
without such feedback show more spheroidal stellar
distributions, older stellar populations, and more total stellar mass than is
observed in nature.  Those cases that do manage to expel enough gas to produce
reasonable stellar mass fractions still fail to produce the correct stellar
kinematics, failing to produce the small bulges characteristic of
so many spiral galaxies.  These models also rely on mechanisms other than
stellar feedback (SMBH heating) or on major numerical contrivances (hydrodynamic
decoupling, etc.).  

Superbubbles offer a way out of this bind. As was shown in \citep{Keller2014},
evaporation in superbubbles naturally mass load winds with temperatures of
$\sim10^7\;K$, a process that is set by self-regulating thermal conduction.
This means that an optimal amount of material is heated above the virial
temperature ($1.3\times10^6\;K$ for a $8\times10^{11}\;M_\odot$ halo).  This
feedback-heated and highly buoyant hot gas migrates out of the disc, cooling
adiabatically while it rises, as is clearly seen in figure~\ref{phase}.   In
fact, as figure~\ref{stellar_fraction} shows, there may even be room to {\it
reduce} the feedback energy below the fiducial $10^{51}\;\rm erg/SNe$, while
still producing reasonable stellar mass fractions.  On top of yielding
physically-realistic galaxies, superbubble feedback is in fact slightly
\textit{less} computationally expensive than blastwave feedback for these
simulations, due in part to its elimination of unphysical high density, high
temperature gas.

\subsection{High-Redshift Outflows Determine Galaxy Properties}

A major failure of other feedback models is 
the production of too many stars, too early.  Superbubble feedback prevents this
by efficiently removing gas from the star forming disc at high redshifts.  The
high mass loadings from $z=2-4$ means more mass is being expelled by
significantly fewer stars.  Mass loadings much larger than unity have been
observed in both Lyman Break Galaxies at high redshift \citep{Pettini2002} and
in local dwarf starburts \citep{Martin2002}.  

Even if a galaxy has the correct stellar mass fraction at $z=0$, forming too
much of your stellar mass at high redshift is problematic for a number of
reasons.  First, it is known from abundance matching as well as the observations
of stellar ages that most stars in $\rm{L^*}$ galaxies are formed fairly late
(e.g. \citet{vanDokkum2013} found $\sim90\%$ of stars in Milky Way like galaxies
formed at $z<2.5$).  Secondly, forming these stars early on means they will be
formed primarily in smaller halos that are subsequently accreted, as well as in
main halo at small radii (as low angular momentum material is accreted first, as
seen in figure~\ref{angular_momentum_netflux}). This results in a stellar
distribution too spheroidal and centrally concentrated, as was seen in many
older simulations, and here in the rotation curves in
figure~\ref{rotation_curve} for the blastwave and no feedback test cases.

Because superbubbles drive strong outflows at high redshift, the galaxy is able
to preferentially remove low angular momentum gas, as indicated by
figure~\ref{angular_momentum_outflow}. This same mechanism  was seen in
simulations of dwarf galaxies by \citet{Brook2011,Brook2012}.  We would suggest
that this is a universal requirement for producing galaxies with low central
concentrations and small bulges.  As \citet{Binney2001} argued, if this gas were
to remain within the disc, it would be ultimately form the massive bulge that is
seen in our blastwave and no feedback cases, and in numerous other past
simulations.  If only a third of the baryons removed in the disc of our fiducial
case were to instead form bulge stars, it would raise the B/T ratio of the disc
to 0.3, making our disc much more bulge dominated than the Milky Way, and
putting our results in conflict with \citet{Allen2006}.  This mechanism (strong,
high-redshift outflows), and the fact that it produces strong suppression of
star formation at high redshift as well as a small stellar bulge, agrees well
with the results of \citet{Sales2012}.  That study of $\sim100$ GIMIC
\citep{Crain2009} galaxies found that those with the smallest B/T ratio also had
the largest fraction of stars formed late in the galaxy's history.  Strong
outflows acting to regulate high redshift star-formation while preserving fuel
for late star formation naturally leads to this situation.

The effectiveness of superbubble feedback may help to prevent the growth of a
massive stellar bulge via a second mechanism as well.  \citet{Fall1980} showed
that the dissipational collapse of hot gas within an extended dark halo can
produce galaxies with thin discs.  However, \citet{Cole2000} showed that the
collisionless processes involved in galaxy mergers can cancel out the net
rotation of a galaxy disc and lead to a spheroidal stellar distribution.  With
superbubble feedback, small dwarves convert very little of their baryonic mass
into stars, allowing them to contribute their gas to the primary halo, which
then collapses dissipationally.  

The heavily mass loaded winds driven by superbubble feedback are not only key to
suppressing high-z star formation, but also to providing enough gas at low
redshift to continue star formation.  The outflows produced by superbubble
feedback can easily escape from the lower-mass progenitors of ${L^*}$ galaxies.
Thus it is able to naturally transition from the violent, wind-driving high
redshift mode to a more quiescent phase as the galactic gravitational potential
and gas surface densities in the disc increase (see the increase in inflow in
figure~\ref{inflow_outflow}.  These factors suppress outflows, and allow star
formation to increase slowly towards $z=0$.

As the disc is assembled, the disc surface density above the 
superbubble increases.  Thus the hot gas must push past
more material and will also entrain more material along with it as shown by
 the clumps seen in figure~\ref{column_density}).  
This slows the outflows compared to those 
seen at high redshift and moving the galaxy
towards a fountain mode.  The gas is kicked to relatively low altitudes and
then rains back down onto the galactic disc.  This effect is probably
sensitive to resolution.  Our small-scale experiments show
 that venting of superbubbles is enhanced by a more porous ISM 
(see \citealt{Keller2014} and also \citealt{Nath2013}).  
Poor resolution suppresses this porosity and increases
numerical dissipation.  Thus better resolution might reduce the puffy
appearance in the HI column density and allow strong galactic
outflows towards $z=0$.  

\subsection{Additional Feedback Mechanisms}

We have shown that, even at moderate resolution, thermal supernova feedback is
sufficient to build a realistic L* galaxy, provided a complete
physical feedback model like that of \citet{Keller2014} is used.    
Thus this work firmly establishes what supernovae and superbubbles can do.

It is still likely that for higher-mass halos $(>10^{12}\; \rm{M_\odot}$),
supernovae alone will not be suffficent to regulate star formation and produce
the drop in star formation efficiency seen in \citet{Moster2013,Behroozi2013}.
We do not include feedback from SMBH, which is likely to be
important for higher-mass galaxies.  

Our resolution limits the formation of dense structures in the ISM.  The lack of
resolved dense gas means that feedback processes involved in disrupting
molecular clouds (UV photodissociation, radiation pressure, etc.) is largely
irrelevant for these simulations.  In fact, since much of the energy from early
stellar feedback is consumed in the disruption of the densest clouds in
simulations such as this, the addition of this energy may be unrealistic,
effectively double counting energy that would have been used to disrupt clouds.
Furthermore, high-resolution studies of individual molecular clouds have found
that ionizing radiation, despite disrupting these clouds, ultimately only
imparts $\lesssim 0.1\%$ of the total radiative luminosity to the gas in the
cloud and the surrounding medium in the form of additional thermal and kinetic
energy \citep{Dale2005, Gendelev2012, Walch2012}).  Thus, applying even as
little as $1\%$ of the radiative luminosity as a source of feedback in a
simulation that does not resolve structure within molecular clouds is likely
massively overestimating the impact on disc scales.

In sufficiently high-resolution simulations it may be necessary to include small
scale feedback mechanisms in order to disrupt clouds before the first SN,
allowing the to explode in a lower density environment since, as
\citet{MacLow1988} showed, the cooling time for superbubbles scales sub-linearly
with the inverse density, $t_R\propto n_0^{-8/11}$.  \citet{Rogers2013} showed
that the disruption of high-density clouds by stellar winds prior to the first
supernovae can allow as much as $99\%$ of the hot SN ejecta to escape into the
surrounding warm ISM, helping to promote the growth of superbubbles that then
vent from the galactic disc.  In the current work, the densities near the
superbubbles are modest and this preprocessing is not required.

\section{Conclusion}

We have shown that supernova feedback alone, with a complete physical
superbubble model, is capable of producing an $\rm{L^*}$ galaxy that falls
within observational constraints.  Superbubbles are a physical mechanism for
producing ab initio galactic winds, that ultimately allow for the formation of a
galaxy with a realistic star formation history and a negligible stellar bulge.

The key results with respect to galaxy formation are as follows:

\begin{itemize}
    \item Strong outflows at high redshift are essential to regulating
        star formation over the total halo history.  The vast majority of stars
        in $\rm{L^*}$ galaxies form after $z\sim2$.  Unless gas is removed from
        galaxies at high redshift by feedback processes, it will rapidly form
        stars, yielding discs that are too massive and red at $z=0$.
    \item Outflows are important for producing the correct disc kinematics and
        preventing the formation of excessive bulges.  Low angular momentum gas
        is accreted first as galaxies form, and the pooling of this gas at the
        center of galaxies can lead to galaxies with sharply peaked rotation
        curves and unrealistically large bulges, often containing the majority
        of stars within the galaxy.  
    \item Galaxies simulated without feedback, or with disabled cooling feedback
        models fail to expel enough of this gas early on, and result in
        bulge-dominated galaxies with unrealistic stellar mass fractions.  The
        superbubble model, on the other hand, produces strong outflows that
        ultimately yield realistic galaxies.
    \item Superbubble feedback naturally yields the sort of outflows that are
        required for $L^*$ galaxies.  The mass-loading \& velocity of
        the winds are set by the hydrodynamics and evaporative mixing, unlike
        other methods where these values are free parameters.
    \item Superbubble feedback produces high-redshift outflows that
        preferentially remove low angular momentum gas, and in so doing,
        prevents the formation of massive bulges and the associated strongly
        peaked rotation curves. It does this without also expelling additional
        high-angular momentum gas, allowing the stellar disc to form while
        arresting the formation of a bulge.
    \item These advantages come without any significant additional computational
        expense, and may in fact be less costly than alternative feedback
        models.
\end{itemize}

Superbubbles are effective up to at least $\rm{L^*}$.   Beyond $\rm{L^*}$, we
anticipate important roles for SMBH feedback or potentially radiation pressure
(see e.g.  \citealt{Hopkins2013}).  In the subsequent MUGS2 series of papers, we
will show how these effects extend to larger and smaller halos forming in a
range of environments.

\section*{Acknowledgements}
The analysis was performed using the pynbody package
(\texttt{http://pynbody.github.io/}, \citep{pynbody}), which was written
by Andrew Pontzen in addition to the authors.  We thank Fabio Governato, Tom
Quinn, Sijing Shen, Greg Stinson, \& Rob Thacker for useful conversations
regarding this paper.  The simulations were performed on the clusters hosted on
\textsc{scinet}, part of ComputeCanada.  We greatly appreciate the contributions
of these computing allocations.  We also thank NSERC for funding supporting this
research.
\bibliographystyle{mnras}
\bibliography{references}

\clearpage

\end{document}